\documentclass[aps,twocolumn,showpacs,amsmath,amssymb,prb]{revtex4}

\usepackage{longtable}
\usepackage{tabularx}

\usepackage[final]{graphicx}
\usepackage{dcolumn}
\usepackage{bm}

\def\figsiz{8cm}

\usepackage{longtable}
\usepackage{braket}

\usepackage[final]{graphicx}
\usepackage{dcolumn}
\usepackage{bm}

\newcommand{\mm}[1]     {\ifmmode {#1} \else{}${#1}$\fi}
\newcommand{\mmm}[1]    {\ifmmode{}#1 \else{}${#1}$\fi}
\newcommand{\beq}[1]{\begin{equation}\label{#1}}
\newcommand{\eeq}{\end{equation}}

\def \cacuni{\mm{\rm Ca_{3}Cu_{3-x}Ni_x(PO_4)_4}}
\def \cacunit{\mm{\rm Ca_{3}CuNi_2(PO_4)_4}}

\def\vec#1{\mm{{\rm\bm{{\mathrm#1}}}}}

\begin{document}
\def\figsiz{\columnwidth}

\title{\large Full propagation-vector star antiferromagnetic order in quantum spin trimer system $\bm\cacunit$}
\author{Vladimir Pomjakushin}
\affiliation{Laboratory for Neutron Scattering, Paul Scherrer
Institut, CH-5232 Villigen PSI, Switzerland}

\date{\today}


\begin{abstract}

We show that the antiferromagnetic structure in the quantum spin trimer system \cacunit\ is based on both arms of propagation vector $\vec{k}$ star $\{[{1\over2},{1\over2},0],[-{1\over2},{1\over2},0]\}$ of the paramagnetic space group $C2/c$. The structure is generated by a symmetric direction of the order parameter of two dimensional irreducible representation of $C2/c$ with one active magnetic mode and corresponds to the Shubnikov magnetic space group $C_a2/c$. We reveal the relation between representation analysis in the propagation vector formalism and Shubnikov symmetry. These types of multi-$\vec{k}$ structures are extremely rarely observed experimentally. To further prove the specific magnetic structure we have performed the calculations of the spin expectation values in the isolated Ni$^{2+}$-Cu$^{2+}$-Ni$^{2+}$ trimer with realistic Hamiltonian. The calculated spin values $\Braket{S_{\rm Ni}}={0.9}$ and $\Braket{S_{\rm Cu}}={0.3}$ are in 10\% accuracy in agreement with the experiment, providing strong complimentary argument in favour of multi-arm magnetic structure. 

\end{abstract}


\maketitle


\section{Introduction}

The low-dimensional magnets have been attracting attention during last years since they show remarkable effects due to the prevalence of quantum physics. In particular, the clusters of spins, such as dimers or trimers with strong intra-cluster interactions and weak inter-cluster ones can show interesting phenomena, for instance field-induced Bose-Einstein condensation of magnons (BEC)~\cite{ruegg03} or quantum magnetisation plateaus \cite{Belik05,MasaCuB}. In addition, the gapped energy spectrum makes these systems interesting for spintronics and also as a quantum computing device material. A potential candidate for one more realisation of the BEC in the spin trimer system \cacuni\ was proposed in Ref.~\cite{furrer07}. It happened to be that the member of this family with $x=2$ is antiferromagnetically (AFM) ordered below $T_N=20$~K with an unusual multi-$\vec{k}$ magnetic structure~\cite{pomjakushin07a}.

In the analysis of magnetic structures on the basis of the neutron diffraction data the most frequent approach is the decomposition of the magnetic representation into irreducible representations (irreps) of the paramagnetic space group  according to Izyumov and Naish  \cite{izyumov91,Kovalev}. In this approach (rep-analysis) only symmetry elements of the space group $G$ that leave the propagation vector ($\vec{k}$-vector) invariant are used to construct normal magnetic modes. This subgroup is called a little group or $\vec{k}$-vector group $G_k$. In general, there are several propagation vectors (arms) forming a so called star. The arms are related by the symmetry elements that are not in $G_k$. In case if the propagation vector star $\{\vec{k}\}$ has several arms the rep-analysis is done in the following way. Practically in all cases it is postulated that the use of only one arm of the star is enough for the description of the experimental data.  The atomic positions which are equivalent in the paramagnetic space group are in general split up into the so called orbits with the atom positions in the different orbits related by the symmetry operators that are not in $G_k$. In the single $\vec{k}$-vector rep-approach the atomic spins on different orbits are not related by symmetry and the rep-analysis alone does not provide symmetry constraints on the spin configurations on different orbits. 

Nowadays there is a growing understanding of the fact that in some cases the use of rep-analysis together with Shubnikov group symmetry or with magnetic superspace (3D+1) groups (e.g., \cite{Perez-Mato2012}) allows one to find a hidden symmetry, which is not evident from the rep-analysis alone. Certain additional constraints on the normal modes obtained from the rep-analysis can be imposed using crystallographic symmetry arguments. This way of analysis is routinely used by the crystallographers in the treatment of the experimental diffraction data on crystal structures, but for some reasons practically not used by physicists in analysis of magnetic diffraction. 
Especially in the cases when the symmetry of the little group is significantly lower than the paramagnetic group, the use of the multi-arm analysis can help to find a high symmetry solution with symmetry related spin configurations on different orbits.

The solution for the magnetic structure \cacunit\ with the use of both arms of the star $\{[{1\over2},{1\over2},0],[-{1\over2},{1\over2},0]\}$ reported in Ref.~\cite{pomjakushin07a} had been found heuristically and it was not clear how unique the solution was and what could be other possibilities.
In the present paper in section II we perform the symmetry analysis using both irreducible representations of propagation vector star and Shubnikov groups. The key difference to the analysis performed in Ref.~\cite{pomjakushin07a} is that here we use irreps of the $\{\vec{k}\}$-star, but not of the single arm.  We determine the respective Shubnikov space group of the multi-arm magnetic structure and show what are the normal modes that enter the magnetic representation decomposition and also discuss all possible configurations within the given symmetry. In addition, there are some advantages to the knowledge of Shubnikov crystallographic symmetry. In particular, this can be useful in the calculations that consider the spin as a quantum object in which case the irreducible co-representations should be used, e.g., the calculations of the magnon excitations as shown in Refs. \cite{DANIEL69,JOSHUA68}. Shubnikov group description also provides strict unified characterisation of the magnetic structure and can be used in various software tools.

To the best of our knowledge the experimentally established cases when the whole star $\{\vec{k}\}$ must be involved in the transition to AFM ordered state are rare. As the examples we can point out on Ref.~\cite{Stewart}, where 4$\vec{k}$-structure was found by analysis of the diffuse scattering in ${\rm Gd_2 Ti_2 O_7}$, Ref.~\cite{Zaharko2003} with 2$\vec{k}$ magnetic structure in ${\rm Ce B_6}$ corroborated by single crystal neutron diffraction and $\mu$SR study, and 3$\vec{k}$-structure in TmAgGe found by powder neutron diffraction \cite{Baran3k}.
The multi-$\vec{k}$ structures with the $\vec{k}$-vectors from the $\{\vec{k}\}$-star should be distinguished from the modulated multi-$\vec{k}$ harmonics structures or the structures with $\vec{k}$-vectors unrelated by symmetry that occur quite often in practice. 

Since the multi-arm structures are rarely reported we believe that some complementary arguments supporting our magnetic structure model would be helpful. To further prove the full star magnetic structure we calculate in section III the spin expectation values in the ground state of the isolated Ni-Cu-Ni trimers with the realistic Hamiltonian parameters as had been determined in \cite{furrer07}. In low dimensional quantum antiferromagnets the truly N\'eel AFM state is not the ground eigenstate resulting in the reduced spin values in case if the AFM ordering occurs. The classical N\'eel state is suppressed by the quantum fluctuations. In the present case of weakly interacting Ni-Cu-Ni trimers one also expects that the spin expectation values will be different from the single ion spin values $s=1$ and $1\over2$ for Ni- and Cu-spins, respectively. Thus, the calculated spin values can be used as an independent verification of the specific type of ordering that results in the specific values of the experimental magnetic moments determined by neutron diffraction. 

\begin{figure}
  \begin{center}
    \includegraphics[width=\figsiz]{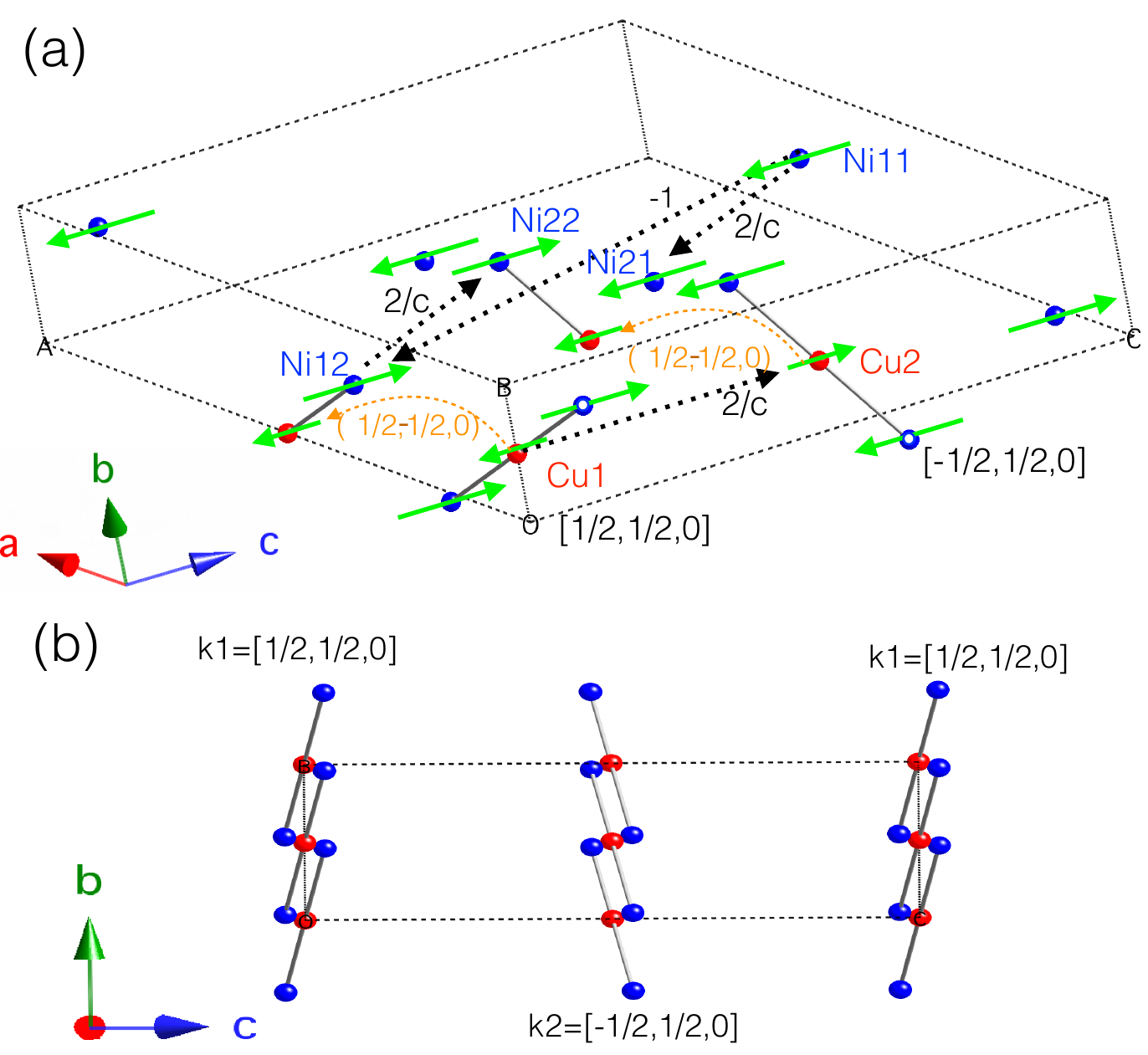}
  \end{center}

\caption{Top (a): The zeroth unit cell of \cacunit\ showing the schematic configuration of the Ni and Cu spins along $c$-axis. Ni and Cu atoms are represented by blue and red circles. The unit cell constants are $a=17.7$~\AA,
    $b=4.8$~\AA, $c=17.8$~\AA, $\beta=123.8^\circ$ ($C2/c$ space
    group). Two of the Ni-spins from the neighbouring cells are shown for better visibility of the trimers as open circles.  Dotted black straight lines indicate the relation between the positions under the symmetry operations: inversion (-$1$) and rotation ($2$) $-x,y,-z+{1\over2}$. The atoms that are not labeled are generated by C-centering translations, as shown for example for Cu1 and Cu2 spins with the orange dashed lines. 
The spins in the middle of the cell along $c$ direction between the dotted lines belong to the orbit 2 (Cu2, Ni21 and Ni22 spins) and have propagation vector $\vec{k}_2=[-{1\over2} {1\over2} 0]$, the other spins belong to the orbit 1 (Cu1, Ni11 and Ni12 spins) and have propagation vector $\vec{k}_1=[{1\over2} {1\over2} 0]$, indicated in the figure. Bottom (b): Projection of the structure along $a$-axis showing trimers on orbits 1 and 2 with propagation vectors $\vec{k}_1$ and $\vec{k}_2$, respectively. The neighbouring trimers running along $b$- and $c$-axes are coupled AFM. The atoms in the trimer are connected by straight solid line.}
  \label{mstr}
\end{figure}

\section{Relation between representation and Shubnikov group descriptions}

In the rep-approach there are two orbits in the magnetic structure: orbit1 with two independent spins Ni11 and Cu1 and orbit2 with Ni21 and Cu2 spins ~\cite{pomjakushin07a}  as shown in  Fig.~\ref{mstr}. The propagation vector space group $G_k$ is triclinic {$C$}-1  [no.2 augmented with centering translations (${1\over2}$,${1\over2},0$)+]. The atom Ni12 on orbit1 is generated by inversion and has to have opposite sign according to irrep $\tau_2$. The returning to zeroth cell translation (1,1,1) reverts the sign back according to formula (\ref{Soft}). All other magnetic moments of the atom $j$ displaced by the translation $\vec{t}$ from its zeroth cell position $\vec{m}_{jl0}$ are obtained by the relation:

\beq{Soft}
\vec{m}_j(\vec{t})=  \sum\limits_{l=1}^2\vec{m}_{jl0}\cos(2\pi \vec{k}_l\vec{t}),
\eeq

\begin{table}
\caption{The magnetic structure parameters in \cacunit\ in (a) propagation vector description (rep-description) in k-vector group {$C-1$} with the atomic positions and magnetic moment components (in $\mu_B$) of the atoms on the orbit 1 Ni11 and Cu1 with propagation vector $\vec{k}_1=[{1\over2} {1\over2} 0]$. The propagation vector, the positions and the magnetic moments of atoms  Ni21 and Cu2 on orbit 2 are generated by the rotation $-x,y,-z+{1\over2}$. Cu1 and Cu2 are in (2c) and (2g) positions, Ni11 and Ni21 are in (4i)-positions, Ni11c and Cu1c are generated by centering translation (0-${1\over2}$-${1\over2}$) and $\vec{k}_1$ using formula (\ref{Soft}); (b) in Shubnikov magnetic space group {$C_a2/c$} (no. 15.91). Ni11 and Ni11c are in (16g)-, Cu1 and Cu1c are in (8a)- and (8b)-positions.}

\label{tab_str}

\begin{center}
\begin{tabular}{l|l|l}
                      &(a)               &(b)                              \\
\hline
$a$, \AA              &17.68079             &33.44705                      \\
$b$, \AA              & 4.80421             &9.608429                     \\
$c$, \AA              &17.79799             &17.79799                     \\
$\beta$, deg         &123.755            &118.477                     \\

\hline
Ni11 $x y z$         &0.62065  0.5353  0.96795  &0.31033 -0.01765 -0.3473              \\
$m_x m_y m_z$        &0.1539 -0.1984 -1.7917      &0.1456 0.1984 1.9466          \\
Cu1  $x y z$         &0 ${1\over2}$ 0           &0 0 0                   \\
$m_x m_y m_z$        &0.3238 -0.1426 -0.3601     &0.3063 0.1426 0.6860         \\
\hline
Ni21 $x y z$         &0.37935 0.5353 0.53205    &             \\
$m_x m_y m_z$        &0.1539 0.1984 -1.7917        &             \\
Cu2  $x y z$         &0 ${1\over2}$ ${1\over2}$ &                   \\
$m_x m_y m_z$        &-0.3238 -0.1426 0.3601       &        \\
\hline
Ni11c $x y z$        &0.12065 0.0353 0.96795    &0.06033 0.23235 -0.8473             \\
$m_x m_y m_z$        &-0.1539 0.1984 1.7917       &-0.1456 -0.19843 -1.9466           \\
Cu1c  $x y z$        &$-{1\over2}$ 0 0          &$-{1\over4}$ ${1\over4}$ -${1\over2}$                   \\
$m_x m_y m_z$        &-0.3238 0.1426 0.3601        &-0.3063 -0.1426 -0.6860        \\

\end{tabular}
\end{center}
\end{table}

where $\vec{t}$ is the centering translation or the unit cell translation. If the structure propagates with one arm of $\vec{k}$-vector star, only one term is left in (\ref{Soft}). 
In this case the spins on orbit1 are not related by symmetry with the spins on orbit2.  The spin configuration with the one arm $\vec{k}_1$, that fits experimental data, has  AFM arrangement of spins in Ni-Cu-Ni trimers on orbit1 and zero values of the spins on the orbit2. Alternatively, there is a second solution that produces equivalent Bragg peak intensities with propagation vector $\vec{k}_2$ and with AFM arrangement of spins on orbit2 and zero spins on orbit1~\cite{pomjakushin07a}. We note that the above configurations generated by $\vec{k}_1$ and $\vec{k}_2$ are different. This is shown in Fig.~1a by dashed orange lines indicating the propagation of Cu1-spin in the trimer on orbit1 and Cu2-spin on orbit2. The trimers displaced by $({1\over2},-{1\over2},0)$ have a ferromagnetic mutual orientation on orbit1, but AFM orientation on orbit2.

\begin{figure}
  \begin{center}
    \includegraphics[width=\figsiz]{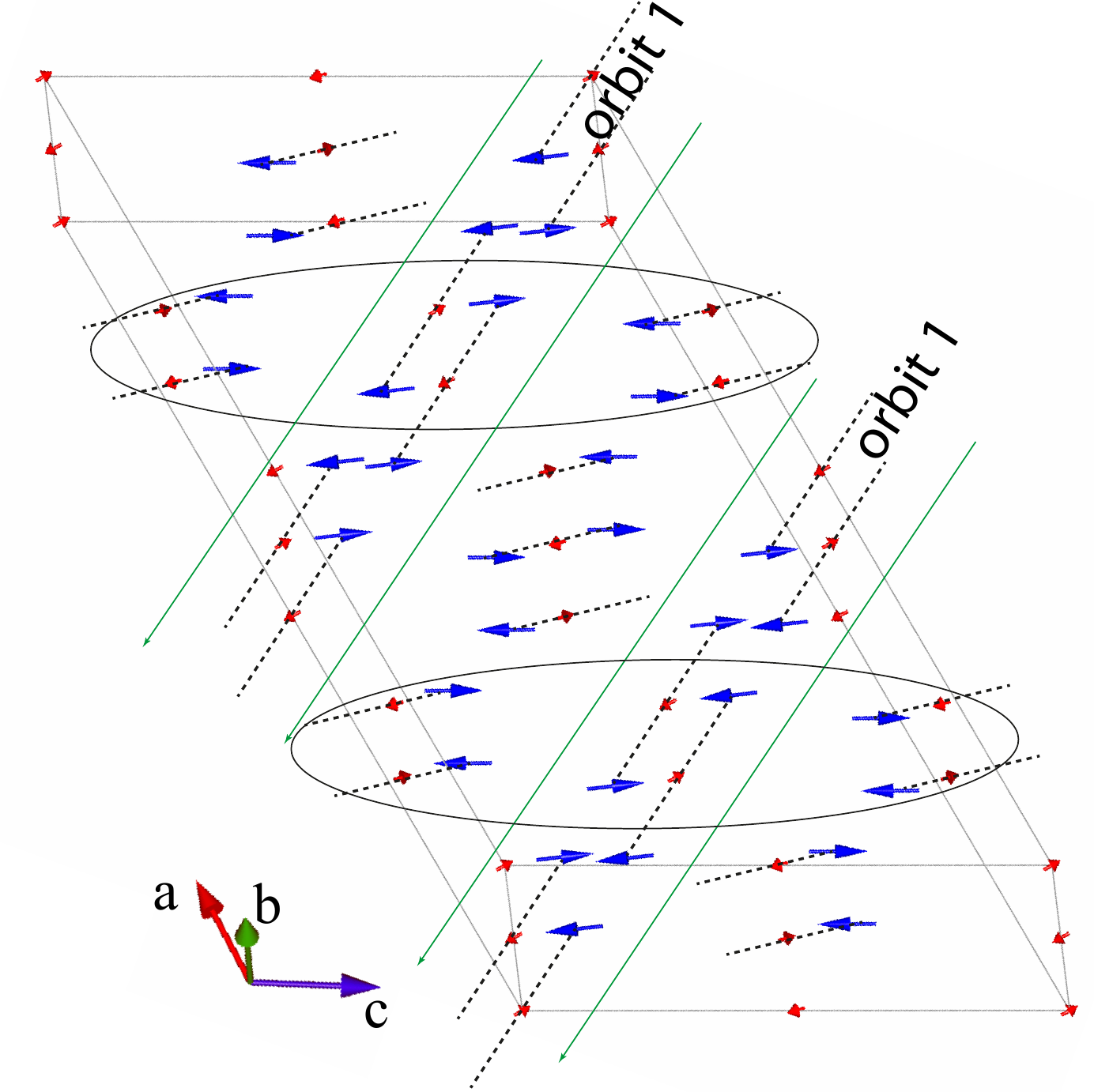}
  \end{center}

\caption{The unit cell of \cacunit\ showing the magnetic structure in Shubnikov group $C_a2/c$ (no. 15.91). Ni and Cu atoms are represented by blue and red circles connected by straight dashed line in the trimers.  The unit cell constants are $a=17.724$~\AA,
    $b=33.4$~\AA, $c=17.8$~\AA, $\beta=118.5^\circ$. The structure is obtained using propagation vector $\vec{k}_1$ for orbit-1 and corresponds to the structure of Fig.~\ref{mstr} and fits the experiment. The structure parameters are listed in Table~\ref{tab_str}. Second possible magnetic structure with equivalent trimers can be constructed using $\vec{k}_2$ for orbit-1. In this structure the spins in the trimers inside the ellipses should be reversed.}
  \label{mstr_sh}
\end{figure}

The symmetry representation analysis  of single-arm structures can be done with the program software tools \cite{Fullprof,MODY,Sarah}.  However, in the present case one has to construct full star structure with symmetry restrictions given by the Shubnikov magnetic space group. This additional symmetry can be revealed by using {\tt ISODISTORT} tool based on {\tt ISOTROPY} software \cite{isot,isod}. The propagation vector $[{1\over2},{1\over2},0]$ (international CDML label of Brillouin zone $V$) in the space group $C2/c$ has two dimensional (2D) irrep denoted as $mV$- \cite{isod}, based on one dimensional irrep $\tau_2$ of the propagation vector group $G_k$. This 2D irrep has three possible directions of oder parameter in the representation space (OPD) that are classified as P1 (a,a), P3 (0,a) and C1 (a,b)  \cite{isot,isod}. The direction P3 is a particular case conventionally used in the rep-analysis when only one arm of the star is used and the spins on the orbits are uncoupled. The direction C1 is the general OPD involving both arms of the star but without symmetry relations between the orbits. Both C1 and P3 directions result in rather low symmetry triclinic group $P_S$-1 (no. 2.7). The most symmetry restrictive direction P1 with the coupled magnetic modes on the two orbits generates highest possible symmetric Shubnikov group $C_a2/c$ (no. 15.91)~\cite{MSG,isod}. The transformation is given by the following matrix: $\vec{A}=2\vec{a}+2\vec{c}$, $\vec{B}=-2\vec{b}$ and $\vec{C}=-\vec{c}$ with the origin shift $\vec{p}={\vec{b}/2}$, where the capital and lowercase letters are the basis vectors for $C_a2/c$ and $C2/c$ space groups, respectively.  In the book \cite{MSG} the Shubnikov group $C_a2/c$ is given in Opechowski-Guccione settings with the symbol $P_C2/c$ (no. 13.8.84). 
To construct the magnetic modes in $C_a2/c$ one should take two atoms on one orbit (it does not matter on which one, let us take orbit1) related by centering translation in the paramagnetic $C2/c$ group and apply the above basis transformation both for spin and for the position. There is no need to transform the atoms on orbit2, because their positions and spins will be generated automatically by the symmetry elements of $C_a2/c$ due to the specific OPD P1.  This transformation results in two independent modes for Cu-spins and two modes for Ni-spins. One mode is generated by the propagation vector $\vec{k}_1$, the second one by the vector $\vec{k}_2$. Since the intra-trimers coupling is dominant one would assume that the spins in the trimers are equally coupled, leaving only two possible solutions. As experimentally found only one mode shown in Fig.~\ref{mstr_sh} fits the experimental data\cite{pomjakushin07a}.  The details of magnetic structure description in both rep-approach and magnetic space group are given in Table \ref{tab_str}. The values of the magnetic moment components in the Table \ref{tab_str} are taken from Ref.~\cite{pomjakushin07a} after the transformation from spherical coordinates to monoclinic axes. There are some differences between Shubnikov and rep-descriptions of magnetic structure. In the Shubnikov description there are two independent atoms of each type: Ni11 and Ni11c, and Cu1 and Cu1c. The rest of atoms, including Ni21 and Cu2, are generated by the symmetry operators of {$C_a2/c$}. In the rep-approach the independent atoms are the ones belonging to the different orbits, i.e. Ni11 and Ni21, and Cu1 and Cu2. The rest of atoms are generated by normal magnetic modes of irrep $\tau_2$ and symmetry operators of {$C$-1} space group with the use of formula (\ref{Soft}).

\section{Energy spectrum and expectation values of the Cu- and Ni-spin in the quantum trimer}

We use dimensionless parametrisation of the Hamiltonian with exchange interaction -1, single ion anisotropy $d$ and magnetic (molecular) field $\vec{h}$. The relation of the dimensionless parameters to the ones in meV used in the paper ~\cite{furrer07} is the following: $d=-D/2J$ and $h=-g\mu_B H_{\rm mf}/2J$. The Hamiltonian with Ni-spin $s={1}$ for $\vec{S}_1$ and $\vec{S}_3$, and Cu-spin $s={1\over2}$ for $\vec{S}_2$ operators reads:

\beq{Ham}
H=\vec{S}_1\vec{S}_2 + \vec{S}_2\vec{S}_3 +  d \sum\limits_{i=1}^3 (S^z_i)^2 - \vec{h} \vec{S},
\eeq
where the operator of total trimer spin is denoted as $\vec{S}=\sum\limits_{i=1}^3\vec{S}_i$. 
The trimer wave function is spanned by 18 basis vectors symbolically denoted as $\chi_{m_1,m_2,m_3}=\Ket{m_1,m_2,m_3}$, where ${m_i}$ are $z$-projections of respective spins in the trimer, which take values $m_1=+1,0,-1$;  $m_2={1\over2},-{1\over2}$ and $m_3=+1,0,-1$.

This section is organised as follows: first we calculate the spin expectation values  $\Braket{\vec{S}_i}$ (the average values of spin operators $\vec{S}_i$) in the trimer for the Hamiltonian without  single ion anisotropy ($d=0$).  In this parameter free model the solution is exact.  Then we calculate $\Braket{\vec{S}_i}$ in the model with the experimentally determined anisotropy $d$ and molecular field $h$. Since the direction of  $\vec{h}$ is not known we vary the angle between the field and anisotropy direction. This gives the variances of of the calculated $\Braket{\vec{S}_i}$ and allows us to make a fair comparison with the experimental values of Ni and Cu spins. We denote the absolute values of the spin expectation values $\Braket{\vec{S}_i}$ by $\Braket{{S}_i}$ (spin sizes).

In  case without single ion anisotropy,  $[H,\vec{S}^2]=0$ and also $[H,S^z]=0$, where $S^z$ is the total spin projection along $\vec{h}$. The total spin quantum number  $S$ is defined by $S(S+1)$ eigenvalue of $\vec{S}^2$ operator with the spin projection $M$ being the eigenvalue of $S^z$.  The commutator $[H,\vec{S}_{13}^2]=0$ holds also for the operator $\vec{S}_{13}^2=(\vec{S}_1+\vec{S}_3)^2$ and we denote its eigenvalues as $S_{13}(S_{13}+1)$. The Hamiltonian is diagonalised with the solution shown in Table 2. In zero magnetic field $\vec{h}=0$ there are 5 degenerate energy levels that are split up into 18 levels by magnetic field as $E_0-M h$. Each energy level can be identified either by the wave function of the trimer in the basis $\chi_{m_1,m_2,m_3}$ or by quantum numbers $S$, $M$, and $S_{13}$ forming a coupled basis. The ground state $E=-3/2(1+h)$ is the one from the quartet $S={3\over2}$ with $M={3\over2}$, assuming that the molecular field is smaller than $h<{5\over2}$. The eigenfunctions with $S={3\over2}$, $M=\pm{3\over2}$ have the form: 

\begin{table}

\caption{Energy spectrum E=$E_0-M h$ of the Hamiltonian (\ref{Ham}) with $d=0$. The spin expectation values along molecular field $\vec{h}$ are $\Braket{S_1}$ for spin-1 at the positions 1 and 3, and $\Braket{S_2}$ for the middle spin-${1\over2}$. The quantum numbers of total trimer spin $S$, its projection on $\vec{h}$ direction $M$, and  ${S}_{13}^2$ are listed. The sum $2\Braket{S_1}+\Braket{S_2}$ is exactly equal to $M$.
}
 \label{hamtab}

\begin{center}
\begin{tabular}{l|l|l|l|l|l}

$E_0$    & $S   $  & $M$    & $S_{13}$ & $\Braket{S_1}$ & $\Braket{S_2}$ \\\hline
-3/2 & 3/2     & $\pm3/2$          &2 & $\pm9/10$ & $\mp3/10$\\
-3/2 & 3/2     & $\pm1/2$          &2 & $\pm3/10$ & $\mp1/10$\\ \hline
-1   & 1/2     &          $\pm1/2$ &1 & $\pm1/3$  & $\mp1/6$\\ \hline
0    & 1/2     &          $\pm1/2$ &0 & $0$       & $\pm1/2$\\ \hline
1/2  & 3/2     &          $\pm3/2$ &1 & $\pm1/2$  & $\pm1/2$\\
1/2  & 3/2     &          $\pm1/2$ &1 & $\pm1/6$  & $\pm1/6$\\\hline
1    & 5/2     &          $\pm5/2$ &2 & $\pm1$    & $\pm1/2$\\
1    & 5/2     &          $\pm3/2$ &2 & $\pm3/5$  & $\pm3/10$\\
1    & 5/2     &          $\pm1/2$ &2 & $\pm1/5$  & $\pm1/10$\\

\end{tabular}
\end{center}
\end{table}

\beq{EF}
{\sqrt{10}\over10}\left( 
\Ket{\pm1,\pm{1\over2},{0}} 
- 2\sqrt{2}\Ket{\pm1,\mp{1\over2},\pm1} 
+ \Ket{0,\pm{1\over2},\pm1}
\right)
\eeq

The spin expectation values in the state (\ref{EF}) are along $z$ axis with the values  equal to $\Braket{S^z_1}=\Braket{S^z_3}=\pm{9\over10}$ for Ni-spins and $\Braket{S_2}=\mp{3\over10}$ for Cu-spin. They are smaller than the single ion spins due to entanglement of the single spin eigenfunctions. In this model there is no one adjustable parameter. It is quite spectacular that this simple model of the isolated trimer in a molecular field $\vec{h}$ is already in a very good agreement with the experimental values $\Braket{S_{\rm Ni}}=0.945(5)$ and $\Braket{S_{\rm Cu}}=0.31(1)$ determined in \cite{pomjakushin07a} (using $g$-factor $g=2$). 

\begin{figure}
  \begin{center}
    \includegraphics[width=\figsiz]{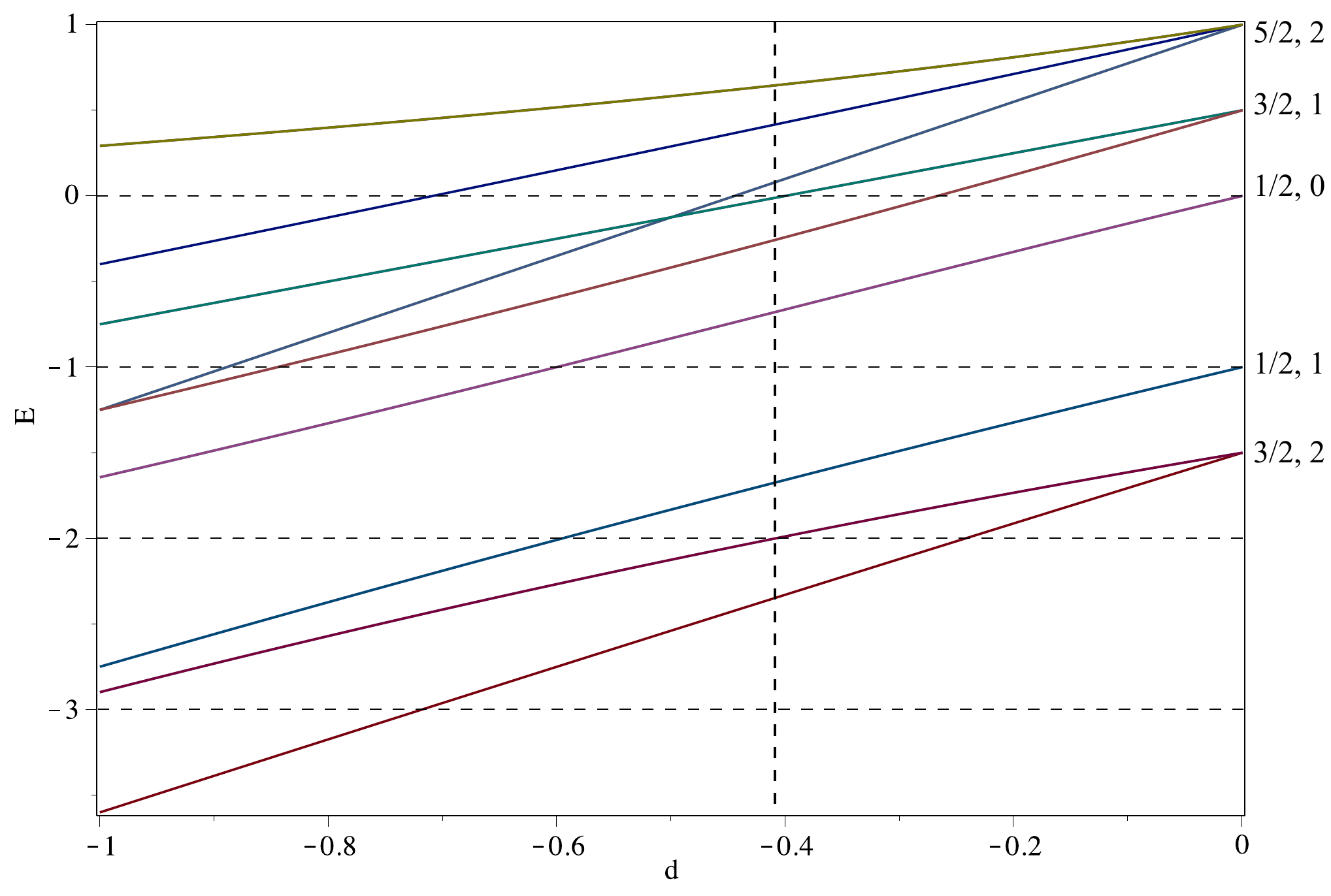}
  \end{center}

  \caption{Energy spectrum E of the Hamiltonian (\ref{Ham}) with $h=0$ as a function of single ion anisotropy $d$. The vertical dashed line indicates experimentally found value $d=-0.41$. For each multiplet with $d=0$ the spin quantum numbers $S$, $S_{13}$ are indicated at the right hand side.}
  \label{Ed}
\end{figure}

\begin{figure}
  \begin{center}
    \includegraphics[width=\figsiz]{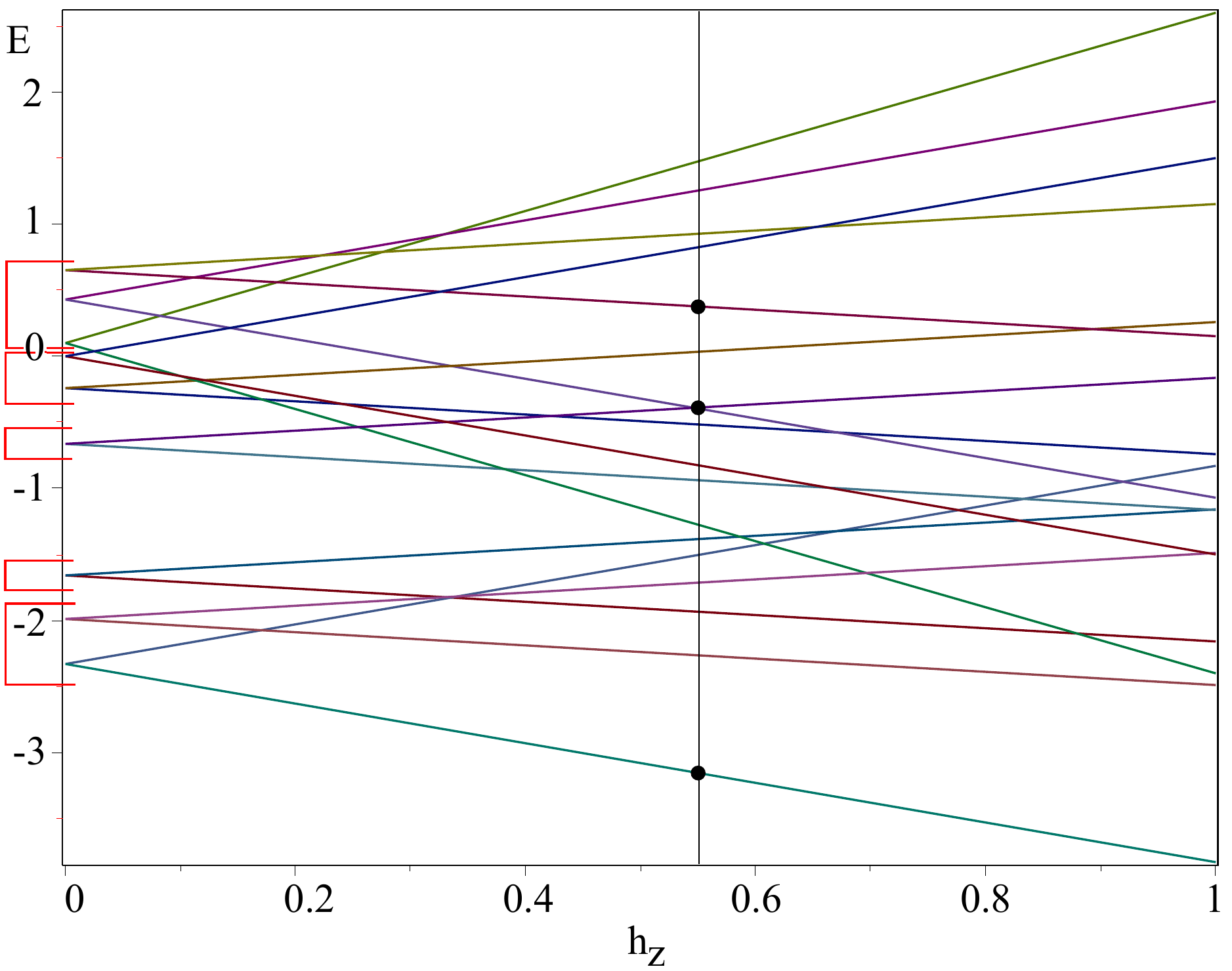}
  \end{center}

  \caption{Energy spectrum E of the Hamiltonian (\ref{Ham}) with $d=-0.4$ as a function of $\vec{h}$ along $z$-axis. The ground state energy is ${1\over4}(7d-\sqrt{4d^2-12d+25}-1)\mp{3\over2}h$ for the lowest at $h=0$ state with $S\simeq{3\over2}, M=\pm{3\over2}$. The square brackets group zero $d$ multiplets that split up into the doublets at $h=0$ with different $z$-projection $M$. The circles indicate two levels from the 5th ``multiplet'' and ground state level that were used to determine the molecular field in \cite{furrer07}. }
  \label{ESz2_h}
\end{figure}

In the case of non-zero single ion anisotropy $d$ with $\vec{h}$ parallel to $z$-axis
the operators $\vec{S}^2$ and $\vec{S}_{13}^2$ do not commute with $H$, and the total spin  $S$ as well as the spin $S_{13}$ are not anymore good quantum numbers. However, due to axial symmetry the $S^z$-projection still commutes with the Hamiltonian. The energy spectrum of (\ref{Ham}) can also be calculated analytically. The ground state quartet $E_0=-3/2$ at $h=0$ with total $S={3\over2}$ splits up into two doublets with spin projection $M=\pm{3\over2}$ for negative $d$ in agreement with \cite{furrer07}. The energy spectrum as a function of anisotropy parameter is shown in Fig.~\ref{Ed}. Since there is no crossing of the energy levels for absolute values of $|d|$ smaller than  $0.41(8)$ (corresponding to the experimental single ion anisotropy of Ni $D_{\rm Ni}=-0.7(1)$~meV), the quantum numbers $S$ and $S_{13}$ at $d=0$ still can be unambiguously used to identify the energy levels.

The magnetic field $h$ along $z$-axis completely removes the degeneracy splitting each doublet in $E_0\mp M h$. Figure \ref{ESz2_h} shows the energy spectrum calculated for the experimental single ion anisotropy of Ni $d=-0.41$. The anisotropy increases the spin expectation values due to the suppression of the terms with $m=0$ in the wave function of the trimer (\ref{EF}). For the given $d$, $\Braket{S_1}=0.92$ and $\Braket{S_2}=0.34$. In the limit of large-$d$ the spins will recover single ion spin sizes $s=1$ and $1\over2$.  We list the first two energy levels $E=-2.35, -2.0$ with $M=\pm{3\over2},\pm{1\over2}$, $\Braket{\vec{S}_{13}^2}=6,5.81$ and $\Braket{\vec{S}^2}=3.77,3.67$, respectively. These values and the other values at $h=0$ shown in the Fig.~\ref{ESz2_h} are in full agreement with the calculations in Ref.~\cite{furrer07} after renormalisation $-(E-E_{GS})2J$, where $E_{GS}$ is ground state energy, $J=-0.85$~meV. 

\begin{figure}
  \begin{center}
    \includegraphics[width=\figsiz]{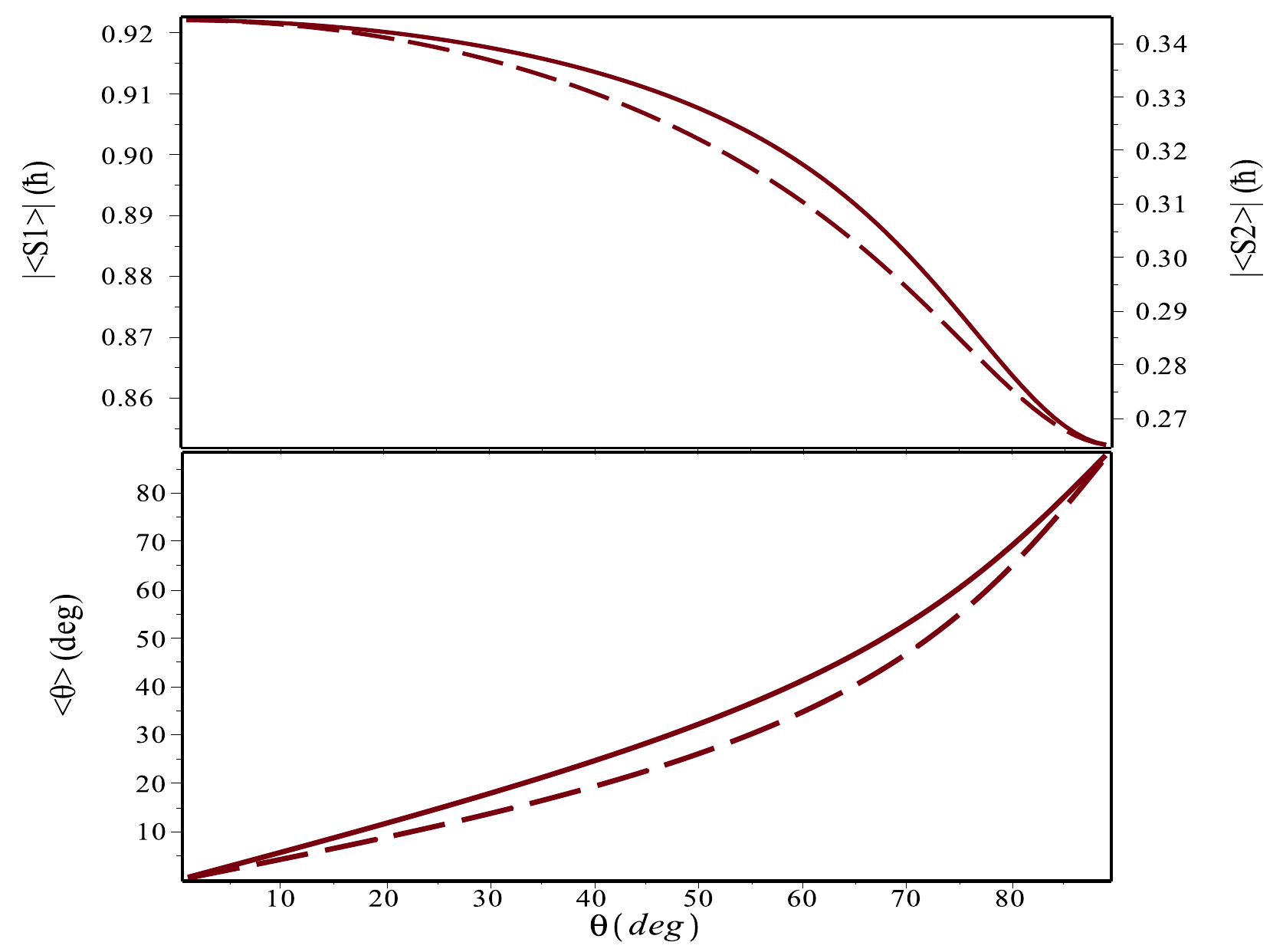}
  \end{center}
  \caption{Absolute values of spin expectation values (upper panel) for Ni $\Braket{S_1}$ and Cu  $\Braket{S_2}$ and their angles $\Braket{\theta}$ with $+z$ axis for Ni and $-z$ axis for Cu  (lower panel) in the ground state of the spin trimer Hamiltonian (\ref{Ham}) with $d=-0.4$ and $h=0.56$ as a function of angle $\theta$ of $\vec{h}$ to $z$-axis. Solid and dashed lines correspond to 1st and 2nd spins in the trimer, respectively.}
  \label{S12th}
\end{figure}

The spin expectation values can change if the molecular field $h$ makes an angle $\theta$ with $z$-axis due to the interplay between the anisotropy $d$ and the molecular field $\vec{h}$. If the field is significantly smaller than $d$, then the spin expectation values will not strongly depend on $\theta$, because small $h$ will not have effect on the wave function, but simply will select the state with $M=+{3\over2}$.  Only for the angles $\theta$ close to 90 degrees the mixing of $\pm M$ states will further reduce the spins $\Braket{S_i}$. 
The molecular field was estimated to be $g\mu_B H_{\rm mf}=0.95(2)$~meV~\cite{furrer07} from the splitting of the energy levels indicated in Fig.~\ref{ESz2_h}. In the above estimation of the field, it was assumed that the splitting does not depend on $d$, which was a fair assumption because the field direction $\vec{h}$ is anyway unknown. This molecular field corresponds to the dimensionless field $h=0.56(7)$. The splitting will not depend on the anisotropy $d$ either if $\vec{h}||z$ or in the limit of large $h$, which seems to be the case. 
The calculations of the energy spectrum as a function of $\theta$ show that the value of $h$ can be underestimated by maximum a factor of 1.8 for the experimentally estimated anisotropy and the field. 
The calculated expectation values of $\Braket{S_i}$ and their angle with $z$-axis $\Braket{\theta_i}$ are shown in Fig. \ref{S12th} as a function of field direction $\theta$ with respect to $z$-axis for the field magnitude $h=0.56$. One can see that the field is large enough and the spins turn  toward the field direction. The spin sizes are slightly further suppressed, but are still in good accordance (about 10\%) with the experimental values. The spins of Ni and Cu in the trimer are not exactly antiparallel, but can make an angle (Fig. \ref{S12th}) down to about $\Braket{\theta_1}-\Braket{\theta_2}\simeq170$ degrees due to the interplay between the anisotropy $d$ and the molecular field $\vec{h}$. This is also in line with the experimental angle $\simeq160$ degrees \cite{pomjakushin07a}. 


\section{Discussion}

In the structural phase transitions it is a regular case when the isotropy subgroup is generated by several arms of the propagation vector~\cite{isot,Hatch_per}, but for the magnetic transitions the multi-arm structures are really an exotics. It might be that in some cases reconsideration of the experimental data with more symmetric multi-arm structure could give similar or even better goodness of fit. 

In the present case one-arm magnetic structure gives physically unreasonable picture. If the spins of all Cu and Ni atoms are allowed to be independent in the data analysis then half of the trimers have zero spin values, as explained in section II and also in Ref.~\cite{pomjakushin07a}. If one forces the spin sizes to be the same (separately for Ni and Cu) in all the trimers then the best possible one-$\vec{k}$ fit to experiment has bad chi-square (goodness of fit). In addition, the Cu-spin gets values that are too big $\Braket{S_{\rm Cu}}=0.6$ and the trimers are non-identical with not necessarily antiparallel Cu and Ni-spins  Ref.~\cite{pomjakushin07a}. The above mentioned one-$\vec{k}$ spin configurations are in contradiction with the calculated spin values and directions in the trimers. The magnetic moments and their orientation obtained from the fit to a very restrictive full star model with Shubnikov symmetry $C_a2/c$ are in amazingly good accordance with the theory as shown in Section III. We find that such good correspondence between experiment and theory is a strong complementary argument in favour of the full star model.
The numerical density functional investigations~ \cite{YareskoDFT} with both inter- and intra-trimer interactions included, but without single ion anisotropy are also in good agreement with the experiment. The calculations give similar magnetic moment value 1.8$\mu_B$ for Ni and slightly larger for Cu 0.8$\mu_B$, and somewhat smaller Cu-Ni intratrimer interaction \cite{YareskoDFT} in comparison with ~\cite{furrer07,pomjakushin07a}. 

Symmetry considerations alone do not restrict the moment sizes to be the same in all the trimers even for the most symmetric direction of order parameter P1 (a,a). In general, the mixing of the modes with different $\vec{k}$-vectors on the same atom will always produce different spin sizes according to (\ref{Soft}). So even the highest Shubnikov symmetry does not force all the trimers to be the same. In the present case the mixing of the mode orbit1+$\vec{k}_1$ with mode orbit1+$\vec{k}_2$ would make the spin sizes inside the ellipses in Fig.~\ref{mstr_sh} different from the other spins. Moreover, if we assume that the spins of Cu and Ni-spin in the trimer propagate with the different arms, then the coupling in the trimers will be both AFM and FM. In the language of Shubnikov group there are two independent Cu-spins in (8a) and (8b) positions and two Ni-spins in (16g) positions, and if one does not couple the spins on different positions, as shown in the Table \ref{tab_str}, then one gets the above mentioned possible spin configurations. In this respect, the rep-approach is very useful because it allows one to choose the appropriate mode based on physical grounds. Namely, the isolated trimer with dominant intracoupling should propagate as a whole object with the same Bloch function given by the propagation vector. This is equivalent to the requirement of having all the trimers identical, leaving only two possible configurations shown in Fig.~\ref{mstr_sh}.

\section{Conclusions}
We have shown that the antiferromagnetic (AFM) structure in the quantum spin trimer system \cacunit\ is based on the full star of propagation vector $[{1\over2},{1\over2},0],[-{1\over2},{1\over2},0]$ of the paramagnetic space group $C2/c$ and corresponds to the Shubnikov magnetic space group $C_a2/c$. The relation between representation analysis in propagation vector $\vec{k}$ formalism and Shubnikov symmetry is examined in details. The unusual multi-$\vec{k}$ (multi-arm) magnetic structure is further supported by the calculations of the spin expectation values  $\Braket{\vec{S}_{\rm Ni}}$ and $\Braket{\vec{S}_{\rm Cu}}$ in the isolated Ni-Cu-Ni trimer with realistic Hamiltonian. 
In the ground state of the trimer in a molecular field the spins are AFM coupled with  $\Braket{S_{\rm Ni}}={9\over10}$ and $\Braket{S_{\rm Cu}}={3\over10}$, being already in a good agreement with the experimental values $\Braket{S_{\rm Ni}}=0.945(5)$ and $\Braket{S_{\rm Cu}}=0.31(1)$ forming the angle about $160$ degrees. Consideration of the realistic single ion anisotropy and molecular field parameters result in the calculated values $\Braket{S_{\rm Ni}}={0.885\pm0.035}$, $\Braket{S_{\rm Cu}}={0.305\pm0.035}$ forming the angle $175\pm5^o$ that within 10\% agree with the experiment, providing strong complimentary argument in favour of multi-arm magnetic structure.

\section*{A{\lowercase{cknowledgements}}}
The computations in this paper were performed by using Maple(TM) analytical software tool~\cite{maple}. We thank Valeri Markushin and Albert Furrer for the discussions and critical reading.

\section*{References}

\providecommand{\newblock}{}


\end{document}